\documentclass[aps,prb,showpacs,amsmath,amssymb,twocolumn]{revtex4-1}
\pdfoutput=1

\usepackage{graphicx}
\usepackage{color}
\usepackage{amsmath}
\usepackage{amssymb}
\usepackage{amsfonts}
\usepackage{bm}
\usepackage{xcolor}
\usepackage{soul}

\DeclareMathOperator{\Tr}{Tr}

\renewcommand{\vec}[1]{\mathbf{#1}}

\begin{document}

\title{Supercurrent Induced Charge-Spin Conversion in Spin-Split Superconductors}

\author{Faluke Aikebaier}
\email[]{faluke.aikebaier@jyu.fi}
\author{Mihail A. Silaev}
\email[]{mikesilaev@gmail.com}
\author{T. T. Heikkil\"a}
\email[]{tero.t.heikkila@jyu.fi}
\affiliation{University of Jyvaskyla, Department of Physics and Nanoscience Center,
P.O. Box 35 (YFL), FI-40014 University of Jyv\"askyl\"a, Finland }

\newcommand{\tmpnote}[1]%
   {\begingroup{\it (FIXME: #1)}\endgroup}
   \newcommand{\comment}[1]%
       {\marginpar{\tiny C: #1}}

\date{\today}

\begin{abstract}
We study spin-polarized quasiparticle transport in a mesoscopic
superconductor with a spin-splitting field in the presence of
co-flowing supercurrent. In such a system, the nonequilibrium state is
characterized by 
charge,
spin, energy and spin energy modes. Here we show that in the presence of both spin splitting and supercurrent, all these modes are mutually coupled. As a result, the supercurrent can convert charge imbalance, that in the presence of spin splitting decays on a relatively short scale, to a long-range spin accumulation decaying only via inelastic scattering. This effect enables coherent charge-spin conversion controllable by a magnetic flux, and it can be detected by studying different symmetry components of the nonlocal conductance signal. 
\end{abstract}


\maketitle

\section{Introduction}
The nonequilibrium states in superconductors can be classified in
terms of energy and charge modes \cite{TinkhamBook,schmid1975}, as
direct implications of the particle-hole formalism in the
BCS theory. In magnetic systems the relevant
nonequilibrium modes are related to the quasiparticle spin. In
spin-split superconductors all these modes need to be considered, and
the quasiparticle diffusion couples pairs of modes
\cite{bergeret2017nonequilibrium,silaev2015long,bobkova2015}. The earlier
description of such spin-resolved modes includes only the direct quasiparticle transport, whereas the effect of supercurrent was not considered. However, a supercurrent flowing along a temperature gradient is known to induce a charge imbalance \cite{schmid1979,clarke1979supercurrent,pethick1980charge,PhysRevLett.43.640}. 
Here we combine these two effects and show how supercurrent couples
all nonequilibrium modes. We show how this leads to  a large
coherently controllable charge-spin conversion induced by supercurrent.
{In particular}, we use the theoretical framework 
\cite{bergeret2017nonequilibrium} based on the quasiclassical
 Keldysh-Usadel formalism for superconductors with a spin-splitting
 field $h$, and consider the presence of a constant
 phase gradient $\nabla \varphi$ in the superconducting order
 parameter. This leads to supercurrent, and shows up
 in the kinetic equations as spectral charge and spin supercurrents. {These coherent supercurrent terms couple spin and charge transport, generating spin from charge injection.} The effect is long-ranged compared to the
spin-relaxation length in the normal state, and becomes very large at
the critical temperature and exchange field. It can be
detected by studying the different symmetry components of the nonlocal conductance. 
 
 The spin-charge conversion {studied here} occurs only under non-equilibrium conditions ($\nabla \mu_{(s)} \neq 0$ or
     $\nabla T_{(s)}\neq 0$) and does not require spin-orbit
     interaction. Therefore it is qualitatively different
     from the direct
     \cite{edelshtein1995,PhysRevB.67.020505,edelstein2005} and
     inverse \cite{edelshtein1989,PhysRevB.92.014509,PhysRevB.70.104521,PhysRevB.76.014522}
     equilibrium magnetoelectric effects proposed for
     noncentrosymmetric superconductors, Josephson junctions \cite{PhysRevLett.101.107005,PhysRevB.92.125443,PhysRevB.94.134506} and superconducting hybrid systems \cite{PhysRevB.95.184518} with spin-orbit coupling. Experimental verification of these {spin-orbit induced} effects is
limited to the recent observations of the anomalous Josephson effect
through a quantum dot \cite{Szombati2016} and Bi$_2$Se$_3$ interlayer \cite{1806.01406,Murani2017}.
{To our knowledge,} the direct magnetoelectric effect, also known as the Edelstein effect, in noncentrosymmetric superconductors have not been observed up to date. In { normal conductors,} such as  
GaAs semiconductors, this effect is known as the inverse spin galvanic effect and has been detected using Faraday rotation.
\cite{PhysRevLett.93.176601}
In contrast, the charge-spin conversion predicted in this work can be measured by purely
electrical probes. Moreover, it is specific to the superconducting
metallic systems and does not rely on the combination of inversion
symmetry breaking and spin-orbit coupling which is usually a tiny effect in such materials.

 \begin{figure}[h!]
 \centerline{$
 \begin{array}{c}
 \includegraphics[width=1.0\linewidth]{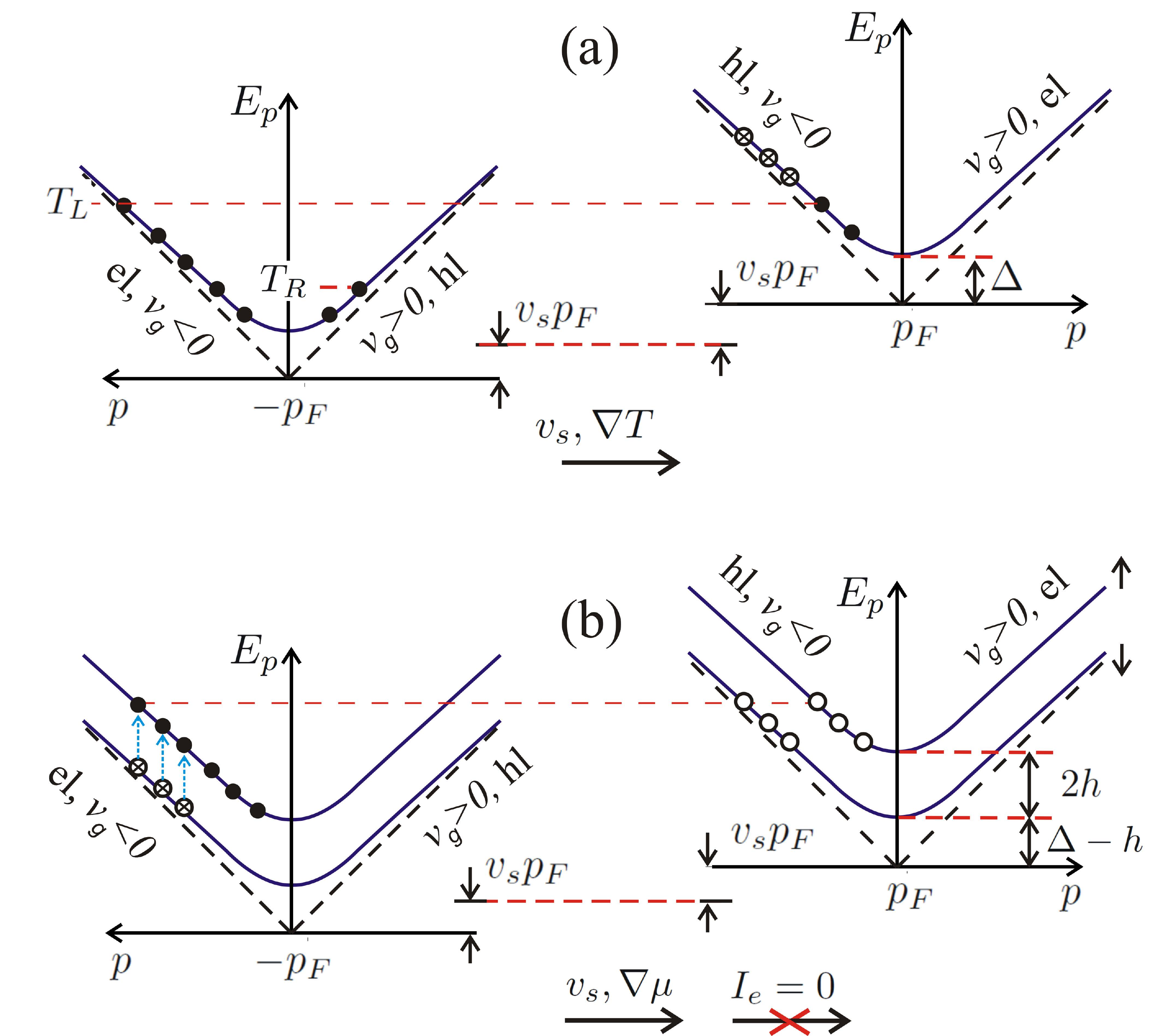}
 \end{array}$}
 \caption{\label{fig:phenom} (Color online) 
 Schematic pictures illustrating the couplings between different types of non-equilibrium states in a superconductor 
 in the presence of 
 the phase gradient driving the condensate to the velocity ${\mathbf v}_s$.
 (a) Generation of charge imbalance by the temperature gradient.
 (b) Generation of spin accumulation by the  charge imbalance gradient $\nabla \mu$ under the 
 restriction that energy current is absent $I_e=0$. 
 Shown in the plots are the quasiparticle electron-like (el) and hole-like (hl) spectral branches in the 
  superconductor in the presence of Doppler shifted energy $\pm p_F v_s$. The filled/open dots show
  the extra occupied/empty states as compared to the equilibrium
  distribution and the dots with crosses show the states which become depopulated due to the Doppler shift.  
  }
 \end{figure}

 \section{Qualitative description of the charge-spin conversion} 
 The 
  supercurrent-generated   coupling between different nonequilibrium states
  can be understood with the
 schematic Fig.~\ref{fig:phenom}, showing the
 spin-split BCS spectrum $E_p+\sigma h  \pm {\bm p}_F{\bm v}_s$ (where $\sigma=\pm 1$ for spin $\uparrow/\downarrow$) 
 for left- and right-moving quasiparticles
  with respect to the condensate velocity direction ${\bm v}_s$.
  The left/right moving states are defined according to their velocities $v_g\equiv \partial E_p/\partial \bm p \gtrless 0$. 
 The balance between the two can be broken either by position
 dependent nonequilibrium modes, or by the presence of a supercurrent
 that induces an energy difference (Doppler shift) $\sim 2 p_F v_s$
 between the states with $p\approx\pm p_F$, where  $p_F$ is the Fermi
 momentum.

       In the absence of spin splitting, $h=0$, the combination of these two effects allows for the creation of charge imbalance  proportional to $v_s \nabla T$. \cite{pethick1980charge,PhysRevLett.43.640,clarke1979supercurrent,schmid1979} 
 This mechanism is illustrated qualitatively in Fig.~\ref{fig:phenom}a. 
    Due to the temperature gradient, left-moving quasiparticles (both electrons e and holes h)
   with velocities $v_e = v_h = -v_g=-v_F \sqrt{E_p^2-\Delta^2}/E_p$ 
   have an excess temperature $T_L$ as compared to that of the right-moving particles $T_R$. 
   From Fig.~\ref{fig:phenom}a one can see that due to the Doppler shift 
   there are more occupied states at the electron branch. This results in the charge imbalance
   controlled by the Doppler shift $p_Fv_s$. {

    Now, let us turn to the system in the presence of velocity $v_s$ 
     and Zeeman splitting $h\neq 0$, shown in Fig.~\ref{fig:phenom}b. 
  }  
 Spin splitting the spectrum provides the possibility for a population difference between spin $\uparrow/\downarrow$ 
 branches. Therefore the supercurrent can couple charge and spin ($\mu_z \propto v_s \nabla \mu$ 
 or $\mu \propto v_s \nabla \mu_z$) as well as excess energy and spin energy 
 ($T_s \propto v_s \nabla T$ or $T \propto v_s \nabla T_s$). Here $\mu_z$ is the
 spin accumulation, and $T_s$ the spin energy accumulation. \cite{bergeret2017nonequilibrium}
 Under general non-equilibrium conditions all these couplings are present. 
 To separate the particular charge-spin conversion effect we must impose certain constraints
 on the distribution function changes due to the supercurrent-induced Doppler shift as 
 in Fig.~\ref{fig:phenom}b. As shown below (Eq.~\eqref{eq:nlg}), these constraints determine the particular
 symmetry components of the non-local conductance as functions of the injector voltage and  polarization 
 of the detector electrode. {For example,} let us assume a charge imbalance gradient  $\nabla\mu\neq 0$ resulting in a 
 larger/smaller number of left-moving electrons/holes in the absence of energy current $I_e$ so that the
 energies of left/right-moving quasiparticles are the same. In the absence of supercurrent these states 
 occupy spin-up/down branches symmetrically yielding no spin accumulation. The Doppler shift results 
 in qualitative changes of  quasiparticle distributions.  
     From Fig.~\ref{fig:phenom}b one can see that in order to have $I_e=0$ without affecting the 
     charge imbalance, the extra energy gained by placing electrons on the Doppler-shifted energy branch
     can be compensated only by utilizing the Zeeman energy and
     shifting some occupied states on the spin-down electron branch to
     the spin-up one (dashed arrows in Fig.~\ref{fig:phenom}). Together with compensating the energy
     difference between left- and right-moving states this shift
     produces a net spin polarization.

 \section{Kinetic theory in the presence of supercurrent and spin splitting}
Below, we quantify the physics described above using the kinetic equations 
\cite{bergeret2017nonequilibrium} based on the quasiclassical
Keldysh-Usadel formalism for superconductors with a spin-splitting field $h$,
to study the spin accumulation generated by the charge
imbalance gradients. For concreteness, we consider the structure in Fig.~\ref{fig:figure2}a. 
A superconducting wire with length $L$ is placed between two
superconducting reservoirs. We assume the presence of a Zeeman
splitting along the wire, either due to a magnetic proximity effect from
a ferromagnetic insulator, or an in-plane magnetic field. A current
is injected in the wire from a normal-metal
injector. A ferromagnetic detector with normal-state conductance
$G_{\rm det}$ and spin polarization $P_{\rm det}$ is placed at 
distance $L_{\rm det}$ from the injector. Variants of this setup were realized for example in Refs.~\onlinecite{quay2013spin,Hubler2012,PhysRevB.87.024517}. 
Here we assume that in addition a homogeneous supercurrent $I_s$ flows along the wire. 
This current can either be driven externally, or it can be induced by a magnetic field in a superconducting loop. 

To study the properties of a mesoscopic superconductor with Zeeman splitting, we start from the Usadel equation
\cite{bergeret2005odd} ($\hbar=k_B=1$)
\begin{equation}\label{eq:UsadelEquation}
D\hat\nabla\left(\check{g}\hat\nabla\check{g} \right)+\left[\check{\Lambda}-\check{\Sigma}_{so}-\check{\Sigma}_{sf}-\check{\Sigma}_{\rm orb}, \check{g} \right]=0,
\end{equation}
where $D$ is the diffusion constant, $\check{g}$ is the quasiclassical Green's function and the covariant 
gradient operator is $\hat\nabla = \nabla - i \bm A [\tau_3, \cdot]$. In the commutator
$\check{\Lambda}=i\epsilon\tau_3-i(\bm{h}\cdot\bm{S})\tau_3-\check{\Delta}$,
$\epsilon$ is the quasiparticle energy, $\bm{h}$ is the spin-splitting
field, $\bm{S}=(\sigma_1,\sigma_2,\sigma_3)$, and the Pauli matrix $\tau_j$
($\sigma_j$) is in Nambu (spin) space. The exact form of the spin-splitting field term, as well as of the pair potential 
$\hat \Delta$ depends on the chosen Nambu spinor. We choose it as

\begin{equation}
\Psi=\left(\psi_{\uparrow}(x),\ \psi_{\downarrow}(x),\ -\psi_{\downarrow}^{\dagger}(x),\ \psi_{\uparrow}^{\dagger}(x) \right)^{T},
\end{equation}
where $T$ denotes a transpose. The advantage of using this spinor is that the Nambu structure has the same form for each spin component. The superconducting pair
potential $\check{\Delta}=\hat{\Delta}\sigma_0$ should be obtained self-consistently (see appendix \ref{sec:selconeq} for details). 
  We denote the Nambu-space matrix  $\hat \Delta (x) = |\Delta| e^{i\varphi (x)\tau_3} \tau_1 $ 
 where
$x$ is the coordinate along the wire. 
Due to supercurrent, the phase $\varphi$ becomes position
dependent. We assume that the quasiparticle currents within the wire
are so small that we can disregard the ensuing position dependence of
$|\Delta|$. The last three terms in the commutator are
$\check{\Sigma}_{so}=(8\tau_{so})^{-1}(\bm{S}\check{g}\bm{S})$,
$\check{\Sigma}_{sf}=(8\tau_{sf})^{-1}(\bm{S}\tau_3\check{g}\tau_3\bm{S})$
and $\check{\Sigma}_{\rm orb}=\tau_{\rm orb}^{-1}\tau_3\check{g}\tau_3$, representing spin and charge imbalance relaxation due to the spin-orbit scattering, exchange interaction with magnetic impurities and orbital magnetic depairing, respectively. The corresponding relaxation rates are  $\tau_{so}^{-1}$, $\tau_{sf}^{-1}$ and $\tau_{\rm orb}^{-1}$. 

We use the real-time Keldysh formalism and describe the quasiclassical Green's function as
\begin{equation}
\check{g}=\left(\begin{array}{cc} \hat{g}^R & \hat{g}^K \\\hat{0} & \hat{g}^A \end{array}\right),
\end{equation}
where each component is a $4{\times}4$ matrix in the Nambu $\otimes$ spin space, $\hat{g}^{R(A)}$ is the
retarded (advanced) Green's function, and $\hat{g}^K$ is the Keldysh Green's function describing the nonequilibrium properties. It can be parameterized in the case of collinear magnetizations by  
$\hat{g}^K=\hat{g}^R\hat{f}-\hat{f}\hat{g}^A$, where the distribution matrix $\hat{f}=f_L+f_T\tau_3+f_{T3}\sigma_3+f_{L3}\sigma_3\tau_3$.

We consider the Eq~(\ref{eq:UsadelEquation}) in the presence of the superconducting current along the wire. 
 Removing the phase of the order parameter by gauge transformation allows us to write Eq.~(\ref{eq:UsadelEquation}) in the
  gauge-invariant form replacing the vector potential by the    condensate momentum 
 $\bm q_s = \nabla\varphi - 2\bm A$.   
 The gradient term in Eq.~(\ref{eq:UsadelEquation}) can be written in the form
 \begin{align} \label{Eq:UsadelGinv}
 & \hat\nabla\cdot(\check{g}\hat\nabla\check{g}) = \nabla\cdot \bm{ \hat I} + \frac{i}{2}[\tau_3, \bm q_s \bm{ \hat I} ] 
 \\ \label{Eq:SpectralCurrent}
  & \bm{ \hat I} = \check g \nabla \check g + \frac{i\bm q_s}{2} \left( \check g \tau_3 \check g - i\tau_3 \right)
 \end{align}
 where $\bm{ \hat I}$ is the matrix spectral current. 
  {We formulate the Keldysh part of this equation in terms of spectral} currents: charge $j_c = {\rm Tr} (\tau_3 \hat I)$, energy $j_e = {\rm Tr} (\tau_0 \hat I)$,
   spin $j_s = {\rm Tr} (\sigma_3 \hat I)$ and spin energy $j_{se} = {\rm Tr} (\sigma_3\tau_3 \hat I)$. 

Kinetic equations derived from Eqs.~(\ref{Eq:UsadelGinv}, \ref{Eq:SpectralCurrent}) 
for these currents can be written in a matrix form 
\begin{equation}\label{eq:KineticEquation}
\nabla\cdot\begin{pmatrix}
j_e\\ j_s\\ j_c\\ j_{se}
\end{pmatrix}=
\begin{pmatrix}
0& 0& 0& 0\\
0& S_{T3}& 0& 0\\
0& 0& R_T& R_{L3}\\
0& 0& R_{L3} & R_T+S_{L3}
\end{pmatrix}
\begin{pmatrix}
f_L\\ f_{T3}\\ f_T\\ f_{L3}
\end{pmatrix},
\end{equation}
where 
\begin{equation}\label{eq:Current1}
\begin{pmatrix}
j_e\\ j_s\\ j_c\\ j_{se}
\end{pmatrix}=
\begin{pmatrix}
D_L\nabla & D_{T3}\nabla & j_E q_s & j_{Es} q_s
\\
D_{T3}\nabla & D_L\nabla & j_{Es}q_s & j_E q_s
\\
j_E q_s & j_{Es} q_s & D_T\nabla & D_{L3}\nabla
\\
j_{Es} q_s & j_E q_s & D_{L3}\nabla & D_T\nabla
\end{pmatrix}
\begin{pmatrix}
f_L\\ f_{T3}\\ f_T\\ f_{L3}
\end{pmatrix}.
\end{equation}
 
The kinetic coefficients $D_{L/T/T3/L3}$, $R_{T/L3}$ and $S_{T3/L3}$
are defined in terms of the components of $\hat g^R$ and $\hat g^A$
(see appendix \ref{sec:KineticCoefficients} and more details in Ref.~\onlinecite{bergeret2017nonequilibrium}). 
The terms $S_{T3/L3}$ are proportional to the total spin relaxation rate in the normal state, 
$\tau_{sn}^{-1}=\tau_{so}^{-1}+\tau_{sf}^{-1}$. The phase gradient provides two additional terms in 
Eq.~(\ref{eq:Current1}): spectral supercurrent $j_E$
\cite{PhysRevB.66.184513} and spin supercurrent
 $j_{Es}=D\Tr[(\hat{g}^R\nabla\hat{g}^R-\hat{g}^A\nabla\hat{g}^A)\sigma_3\tau_3]/(8q_s)$.

In equilibrium $f_L=\tanh(\epsilon/2T)\equiv n_0$ and other modes
are absent. Then the spectral current terms yield non-zero charge
supercurrent $I_s$ and spin-energy  current $I_{se}$ as 
\begin{align}
& I_s=G_{\xi_0} \xi_0 q_s  \int_{-\infty}^{\infty} d\epsilon j_E\tanh\left({\frac{\epsilon} {2T}}\right) 
\\
& I_{se}=G_{\xi_0} \xi_0 q_s  \int_{-\infty}^{\infty} d\epsilon \epsilon j_{Es}\tanh\left({\frac{\epsilon} {2T}}\right),\label{eq:securrent}
\end{align}
where $G_{\xi_0}=e^2 D \nu_F A/\xi_0$ is the normal-state conductance of the
wire of one superconducting coherence length $\xi_0=\sqrt{D/\Delta}$, with normal-state
density of states $\nu_F$ and cross section $A$. We assume that the phase gradient is small so that $I_s$ is much below the critical current of the wire.

\begin{figure}[h!]
 \centerline{$
 \begin{array}{c}
 \includegraphics[width=0.5\linewidth]{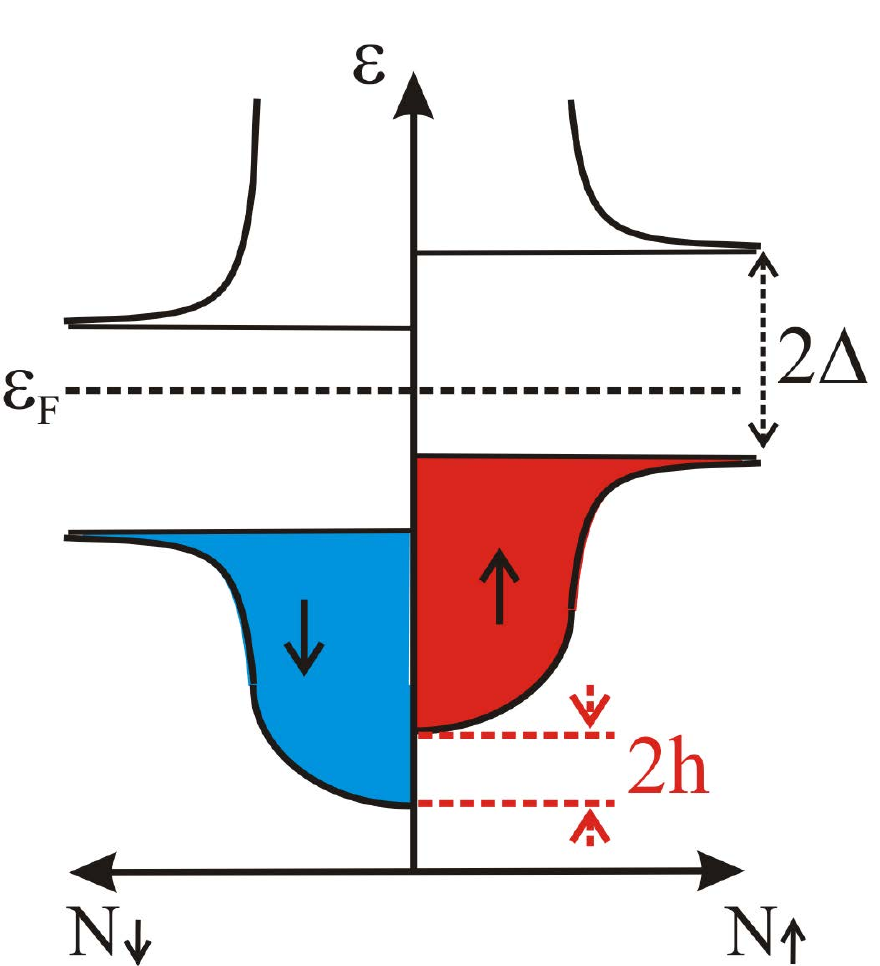}
 \end{array}$}
 \caption{\label{fig:SpinEnergy} 
 Schematic picture illustrating the non-zero spin energy in the
 ground state of a spin-singlet superconductor with spin splitting. $N_{\uparrow, \downarrow}(\varepsilon)$ are the spin-up/down densities of states as functions of the energy $\varepsilon$. The relative Zeeman shift of the electronic bands is $2h$. The case of $T=0$ is shown, so that all states below the Fermi level $\varepsilon_F$ are occupied. } 
    
 \end{figure}

The equilibrium spin-energy current, Eq.~\eqref{eq:securrent}, arises due to the 
modification of the superconducting ground state in the presence of an
exchange field. This is illustrated schematically in
Fig.~\ref{fig:SpinEnergy}, which shows the occupied energy states in
spin-up and spin-down subbands in a superconductor with a spin-splitting field. Here one can see that 
there is a relative energy shift between the spin-up/down subbands. The overall energy difference between these states yields the
non-vanishing spin energy density
$\epsilon_\uparrow - \epsilon_\downarrow = h N_0$, where $N_0$ is the total electron density. 
Since all these particles are in the condensed state, the collective motion of the condensate
results in the coherent spin-energy flow $I_{se} = v_s N_0 h$. 
However, such an equilibrium spin-energy current is not directly observable and can be 
revealed through its coupling to the superconducting current and charge imbalance discussed below.
 
Out of equilibrium, the matrix in Eq.~(\ref{eq:Current1}) couples the four modes together. The diffusion coefficients  
$D_{T3/L3}\neq0 $ for $h\neq 0$
combine pairwise $f_T$ and $f_{L3}$ (charge and spin energy) modes as well as $f_L$ and  $f_{T3}$ (energy and spin) modes \cite{silaev2015long,bobkova2015}. 
An additional coupling between $f_L$ and $f_T$ modes is introduced by $j_E$, mixing charge imbalance with energy. 
This coupling leads to the supercurrent-induced charge imbalance in the presence of a
temperature gradient \cite{pethick1980charge,PhysRevLett.43.640,clarke1979supercurrent}. 
The presence of $h$ and $j_E$ combines these
two effects together in Eq.~(\ref{eq:Current1})
and allows for the conversion between charge imbalance and spin accumulation. In the next section we study the observable consequences of this conversion.

\section{Spin-charge conversion in a non-local spin valve}

\begin{figure}
 \centering
 \includegraphics[width=\columnwidth]{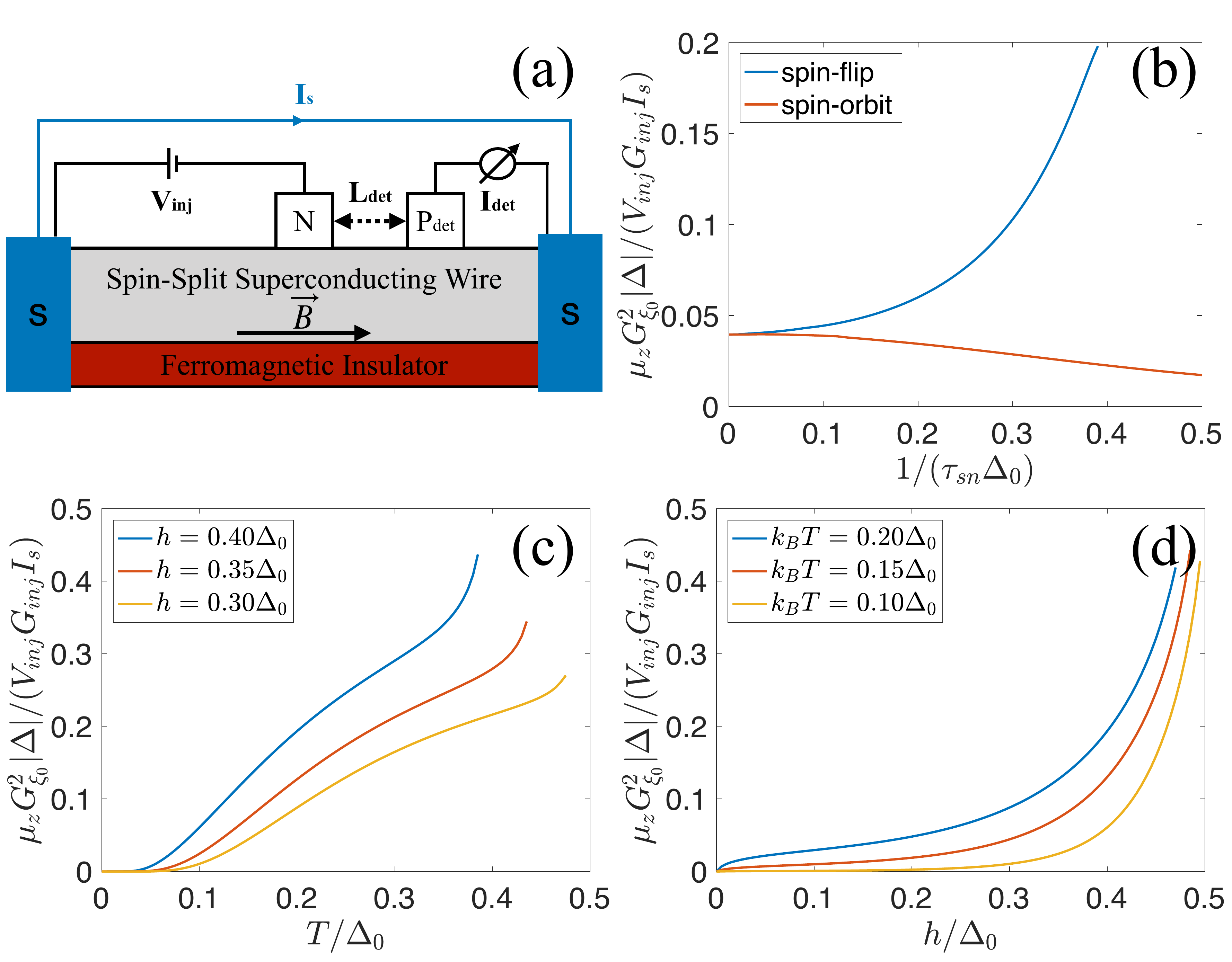}
 \caption{\small (a) Schematic view of the setup. Here the
 spin-splitting field is induced from either the ferromagnetic
 insulator or external magnetic field $\bf{B}$. 
 (b-d) Spin accumulation as a function of parameters $\mu_z=\mu_z(h,T,\tau_{sn})$ at the detector 
 position $L_{\rm det}=L/8$ in the linear response regime (small $V_{\rm inj}$).
 (b) The dependence on the spin relaxation rate for $k_B T=0.15\Delta_0$ and $h=0.3\Delta_0$. 
 (c) Temperature and (d) spin-splitting field dependence. 
 The orbital depairing rate is $\tau_{\rm orb}^{-1}=0.176h^2/\Delta_0$. Here we normalize the induced spin signal by the supercurrent amplitude $I_s$.  }
 \label{fig:figure2}
 \end{figure}

Kinetic theory developed in the previous section can be applied to predict the 
experimentally measurable consequence of charge-spin conversion effect in the 
non-local spin valve setup shown in 
Fig.\ref{fig:figure2}a. It consists of a superconducting wire with externally induced {supercurrent},
injector electrode attached at $x=0$ and ferromagnetic detector electrode attached at some distance
$x=L_D$. The overall length of the wire $L$ is fixed by the boundary conditions which require all non-equilibrium 
modes to vanish at $x=\pm L/2$.

 Consider a non-ferromagnetic injector electrode attached at $x= 0$. We describe the injection 
 of matrix quasiparticle current using the boundary conditions at the tunnelling interface \cite{kuprianov1988influence}
 extended to the spin-dependent case  
 \cite{bergeret2012electronic}

  \begin{equation} \label{Eq:bcDistrFunc}
 \left( \begin{array}{cccc}
  \left[j_c\right] \\ \left[j_{e}\right] \\ \left[j_s\right] \\ \left[j_{se}\right]
  \end{array} \right)  =
    \left(
    \begin{array}{cccc}
    N_+ & PN_- & PN_+ & N_-  \\
    PN_- & N_+ & N_- & PN_+ \\
    PN_+ & N_- & N_+ & PN_- \\
    N_- & PN_+ & PN_- & N_+ \\
   \end{array}
   \right)
   \left(\begin{array}{cccc}
   \left[ f_T\right] \\  \left[ f_L \right]  \\ \left[f_{T3}\right] \\ \left[ f_{L3} \right]
   \end{array} \right) \; .
   \end{equation} 
   Here
 the {left hand side} of Eq.~(\ref{Eq:bcDistrFunc}) contains the differences between currents 
   in the superconducting wire on the left and on the right from the injector, 
    $[j_k] = [j_k(x=+0) - j_k(x=-0)]/\kappa_I$, where $k=T,L,T3,L3$
    and 
      $\kappa_I=G_{\rm inj}/(G_{L}L)$ is the injector transparency defined by the ratio of the
   normal-state conductance $G_{\rm inj}$ of the injector and the conductance $G_L L$ of the wire per unit length.     
    
     {The right hand side of Eq.~(\ref{Eq:bcDistrFunc}) contains the  differences} of the distribution function components
   $[f]_k = f^{(S)}_k - f^{(N)}_k  $  between the superconductor and normal-metal electrodes. 
    The response matrix is here described by the spin polarization $P$ and the energy-symmetric and 
    energy-antisymmetric parts of the density of states, $N_+ = {\rm Tr}\; {\rm Re}(\tau_3 \hat g^R)$ 
    and $N_- = {\rm Tr}\; {\rm Re}(\sigma_3\tau_3  \hat g^R)$. 
    In our particular case the normal-metal 
    injector is characterized by the Fermi distribution function shifted by the applied bias voltage 
    $V_{inj}$. Therefore  
    we have $[f_L]= f_L-n_+$, $[f_{T}]= (f_T-n_-)$, $[ f_{T3}]= f_{T3}$ and 
    $[ f_{L3}]= f_{L3}$,
    where 
    $n_{\pm}=[n_0(\epsilon+V_{inj})\pm n_0(\epsilon-V_{inj})]/2$.

 The solutions of Eqs.~(\ref{eq:KineticEquation},\ref{Eq:bcDistrFunc})
 can be used for calculating the tunnelling current $I_{det}$ measured by a spin-polarized  detector \cite{silaev2015long}
 with spin-filtering efficiency $P_{det}$
 \begin{align} \label{Eq:Idet}
 & I_{\rm det}=G_{\rm det}(\mu+P_{\rm det}\mu_z)
  \\ \label{Eq:ChPot0}
 & \mu = \frac{1}{2}\int_{-\infty}^{\infty} d\varepsilon ( N_+ f_T+ N_-  f_{L3}) \\\label{Eq:ChPotZ}
 & \mu_z = \frac{1}{2}\int_{-\infty}^{\infty} d\varepsilon [ N_+ f_{T3}+ N_-(f_{L}-f_{\rm eq})].
\end{align}

 The contributions from the different nonequilibrium modes to $\mu$ and $\mu_z$ can be read off from the different symmetry components of $I_{\rm det}$ with respect to the injection voltage $V_{\rm inj}$ and the detector polarization $P_{\rm det}$.
The non-spin-polarized injector generates charge $f_{T}$ and energy $f_{L}$ modes \cite{PhysRevLett.28.1366}, which are odd and even in the injection voltage, respectively.
In spin-split superconductors the energy mode is coupled to the spin accumulation producing a long-range spin signal with the symmetry \citep{silaev2015long}
$\mu_z (V_{\rm inj}) = \mu_z (-V_{\rm inj})$. The supercurrent converts part of the charge imbalance to long-range spin accumulation with the opposite symmetry $\mu_z (V_{\rm inj}) = - \mu_z (-V_{\rm inj})$.

  Below we concentrate on the details of this mechanism.

  At first,  we solve the 
 kinetic equations using a perturbation expansion in the small parameter $\xi_0 q_s$ where
 $\xi_0=\sqrt{D/\Delta}$ is the coherence length. 
  For simplicity, we disregard inelastic scattering that would add an energy-non-local term in Eq.~\eqref{eq:KineticEquation}, 
and rather assume that $f_L=n_0$ at the ends of the wire. This mimics the typical experimental situation where the wire 
ends in wide electrodes, often at a distance small compared to the inelastic scattering length. In this case the solution of $f_L$ includes a linear component. The solution of $f_{T3}$, however, is determined by the strength of spin relaxation. 
  This calculation is detailed in appendix \ref{sec:PerTheo}. 

  When $q_s=0$ 
 we find $f_T$ and $f_{L3}$ modes generating the charge imbalance $\mu$.  
 For $q_s\neq 0$ [see Eq.~\eqref{eq:Current1}]  these solutions provide sources for the $f_L$ and $f_{T3}$ 
 modes generating the spin accumulation $\mu_z$ 
 in accordance with the qualitative mechanism illustrated in Fig.~\ref{fig:phenom}b. 
 This generation  takes place close to the injectors, before the charge imbalance relaxes due
to the presence of an exchange field and depairing
\cite{bergeret2017nonequilibrium,hubler2010}
(blue lines in Fig.~\ref{fig:figure3}a). 
However, $\mu_z$ has a long-range part associated with the contribution of $f_L$, 
 which consists of two qualitatively different parts discussed below.

  First, even in the absence of the supercurrent there exists a long-range contribution 
  related to the already known  heating effect \cite{silaev2015long} given by 
   \begin{equation} \label{Eq:LongRange2}
 f^{heat}_{L} (x)= \alpha_{heat} ( |x| -L) 
\end{equation} 
 where $\alpha_{heat} = N_+ n_+ /D_L$.
 Besides that the long-range contribution excited due to the supercurrent is given approximatively by 
 \begin{equation} \label{Eq:LongRange1}
 f^{super}_L = \alpha_{super} ({\rm sign}(x) - x/L) 
\end{equation}  
 The amplitude $\alpha_{super}$ depends on the strength of relaxation described by $R_{T/L3}$ and $S_{T3/L3}$ in Eq.~\eqref{eq:KineticEquation}.  
  
   Note that the spatial structures of  (\ref{Eq:LongRange2}) and (\ref{Eq:LongRange1}) are different 
    because $f^{heat}_{L} (x)$ is an even function and $f^{super}_{L} (x)$ is an odd function of $x$, see Fig.~\ref{fig:figure3}a.
 Besides that, the amplitude of supercurrent-induced part is an odd function of the 
 injector voltage $\alpha_{super} (V_{inj})= - \alpha_{super} (-V_{inj})$. Therefore it exists already in the linear regime
 whereas the heating (\ref{Eq:LongRange2}) is a nonlinear effect since $
 \alpha_{heat}(V_{inj})= \alpha_{heat}(-V_{inj})$. Besides  that, as one can see from  Eq.~(\ref{Eq:LongRange2}),
 the heating contribution grows linearly with the wire length $L$ while the supercurrent-related part (\ref{Eq:LongRange1})
 does not depend on the length $L$ at distances $|x|\ll L$.

To gain further insight, we first study the spin accumulation using a numerical solution of the 
kinetic equations. 
  In Figs.~\ref{fig:figure2}b-d and ~\ref{fig:figure3}a,b, we show the dependencies of the spin accumulation 
 on various parameters $\mu_z=\mu_z(h,T,\tau_{sn},V_{\rm inj},x) $ obtained from the
 numerical solutions of Eqs.~(\ref{eq:KineticEquation},\ref{eq:Current1}). 
 Note that from this plot it is clear that the effect exists entirely due to the modification of quasiparticle spectrum by the spin splitting: As shown in Figs.~\ref{fig:figure2}c,d the spin signal $\mu_z$ disappears both for $h\to 0$ when there is no spin splitting and for $T\to 0$ when there are no quasiparticles. At the same time, Fig.~\ref{fig:figure2}b shows that the effect survives in the absence of spin-orbit or spin-flip scattering, i.e., for $\tau_{sn}\to \infty$.  
  Below we study in more detail the influence of spin relaxation on the 
 behaviour of different contributions to the spin accumulation.

\subsection{Case without spin relaxation ($S_{T3,L3} = 0$)} 

The discussed mechanism of spin-charge conversion does not require any non-conservation of spin. This makes a qualitative distinction with previously discussed direct and inverse Edelstein effects which rely on the spin-orbit interaction.\cite{edelshtein1995,PhysRevB.67.020505,edelstein2005,edelshtein1989,PhysRevB.92.014509,PhysRevB.70.104521,PhysRevB.76.014522} 
{In the absence of spin relaxation, }
$f_{T3} \propto x$ is also a long-range mode similar to the longitudinal one which in the absence of inelastic scattering is long-ranged, see Eqs.~(\ref{Eq:LongRange2},\ref{Eq:LongRange1}).
The combination of $f_{T3}$ and $f_L$ then yields 
(see details in appendix \ref{sec:PerTheo})
\begin{equation}\label{eq:SpAccIntsnInf}
\mu_z=\xi_0\partial_x\varphi\frac{G_{\rm inj}}{G_{\xi_0}}\int_0^{\infty}d\epsilon
n_-(\epsilon;V_{\rm inj})
\sum_{\sigma=\uparrow,\downarrow}\frac{\sigma N_\sigma^2j_s^\sigma}{4D_L^\sigma R_T^\sigma}u_0(x).
\end{equation}
Here $u_0(x)=-u_0(-x)$ is a function that decays linearly from unity
close to the injector ($x=0$) to zero at the reservoirs and $n_-=[n_0(\epsilon+V_{\rm
  inj})-n_0(\epsilon-V_{\rm inj})]/2$. Equation (\ref{eq:SpAccIntsnInf})
describes the region $|x|>\lambda_{cr}$, where $\lambda_{cr}$ is the
charge relaxation length. 
Here $N_{\uparrow/\downarrow}$ are spin-up/down density of states, 
$D_L^{\uparrow/\downarrow}=D_L \pm D_{T3}$, 
$R_T^{\uparrow/\downarrow}=R_T\pm R_{L3}$, and
$j_s^{\uparrow/\downarrow}=j_E\pm j_{Es}$. Moreover, $G_{\rm inj}$ and
$G_{\xi_0}$ are the normal-state conductances of the injector and of
a wire with length $\xi_0$, respectively. The integrand in
Eq.~\eqref{eq:SpAccIntsnInf} is peaked at $\epsilon \approx \Delta \pm
h$ due to the BCS divergence in $N_\sigma$, $j_s^\sigma$ and
$R_T^\sigma$. This divergence can be cut off by the depairing
parameter \cite{dynes84} $\Gamma$  so that for $\epsilon = \Delta +
\sigma h$, $N_\sigma \approx
\gamma_\sigma^{-1/2}/\sqrt{2}$, $j_S^\sigma \approx \gamma_\sigma^{-1}/2$ and
$R_\sigma \approx \gamma_\sigma^{-1/2}/2$ with $\gamma_\sigma =
\Gamma/(\Delta+\sigma h)$. Therefore the integrand scales as
$(8\gamma_\sigma)^{-3/2}$, whereas the width of the peak is $\propto
\Gamma$. Overall, this means a diverging integral scaling like $\sim
\Gamma^{-1/2}$. Similar divergence was found in Ref.~\onlinecite{schmid1979}
for the supercurrent induced charge imbalance in the absence of spin splitting. 

 In practice, the relevant depairing mechanism in the presence of spin
splitting and supercurrent is the orbital depairing due to the combined effect of the supercurrent itself and of an
in-plane magnetic field $\vec{B}$ on the spectrum of the superconductor
\cite{deGennes:566105,belzig96,anthore2003}, with rate 
 $\tau_{\rm orb}^{-1}=D \Delta (\partial_x \varphi)^2/(2)+De^2 B^2 d^2/6$ for a film with
thickness $d$. 
It does not relax the spin, but affects the spectral properties of the
superconductor by reshaping the singularities in the spectral
quantities \cite{bergeret2017nonequilibrium}. We can hence use
$\tau_{\rm orb}^{-1}$ instead of $\Gamma$ to cut the divergence, and see that for very large phase
gradients, $\mu_z$ becomes independent of $\partial_x \varphi$.

According to Eq.~(\ref{eq:SpAccIntsnInf}) the difference of the
quantity $N_0^2j_s/(D_LR_T)$ for spin up and down species describes
the charge-spin conversion. We find that the charge imbalance
is proportional to the energy integral of $N_0^2/R_T$,
averaged over spin. The charge is then converted to spin at a rate
$\propto j_s/D_L$.  
The temperature and exchange field dependence of $\mu_z$ are given in
Figs.~\ref{fig:figure2}c and d, respectively. We can see that the
linear-response $\mu_z\rightarrow0$ as $T\rightarrow0$, which reflects
the freezing of the quasiparticle population (Fig.~\ref{fig:figure2}c). However, this can be
circumvented by considering response at $V_{\rm inj}\sim\Delta$ as shown below. At the
superconducting critical temperature $T_c$, the ratio $\mu_z/I_s$
diverges similarly to the supercurrent induced charge imbalance in the
presence of a temperature gradient
\cite{clarke1979supercurrent,pethick1980charge}. Since $T_c$ is lower
for a higher exchange field, this divergence happens at a lower
temperature in a higher exchange field. For a fixed temperature, the divergence of $\mu_z$ also happens at a critical exchange field (Fig.~\ref{fig:figure2}d) where superconductivity is suppressed \cite{Chandrasekhar1962,Clogston1962}.

\subsection{Effect of spin relaxation}
Spin-flip and spin-orbit relaxation
affect both spectral and nonquilibrium properties of the superconductor. For the spectral properties,
spin-flip relaxation breaks the time-reversal symmetry and suppresses
the superconducting pair potential and critical temperature, while
spin-orbit scattering reduces the effect of the exchange field without
suppressing the pair potential \cite{bergeret2017nonequilibrium}. Both
spin-flip and spin-orbit scattering also lead to the relaxation of 
$f_{T3}$ [terms $S_{T3/L3}$ in
Eq.~\eqref{eq:KineticEquation}]. For strong spin relaxation, the
contribution to $\mu_z$ thus results only from $f_L$, and
decays only via inelastic scattering. In this case (see details in appendix \ref{sec:PerTheo})
\begin{equation}\label{eq:muzInStroRelax}
\mu_z= \xi_0\partial_x\varphi_0\frac{G_{\rm inj}}{G_{\xi_0}}
\int_0^{\infty}d\epsilon n_-(\epsilon;V_{\rm inj}) \frac{\left(N_{\uparrow}^2-N_{\downarrow}^2\right)j_E}{4R_TD_L}u_1(x).
\end{equation}
Here the linear function $u_1(x)=-u_1(-x) \approx u_0(x)$ for
$|x|>\lambda_{cr}$. However, the effects of spin-flip/spin-orbit
scattering on the spectral functions also affect the resulting $\mu_z$. The effect depends strongly on the type of scattering.

 \begin{figure}
    \centering
   \includegraphics[width=\columnwidth]{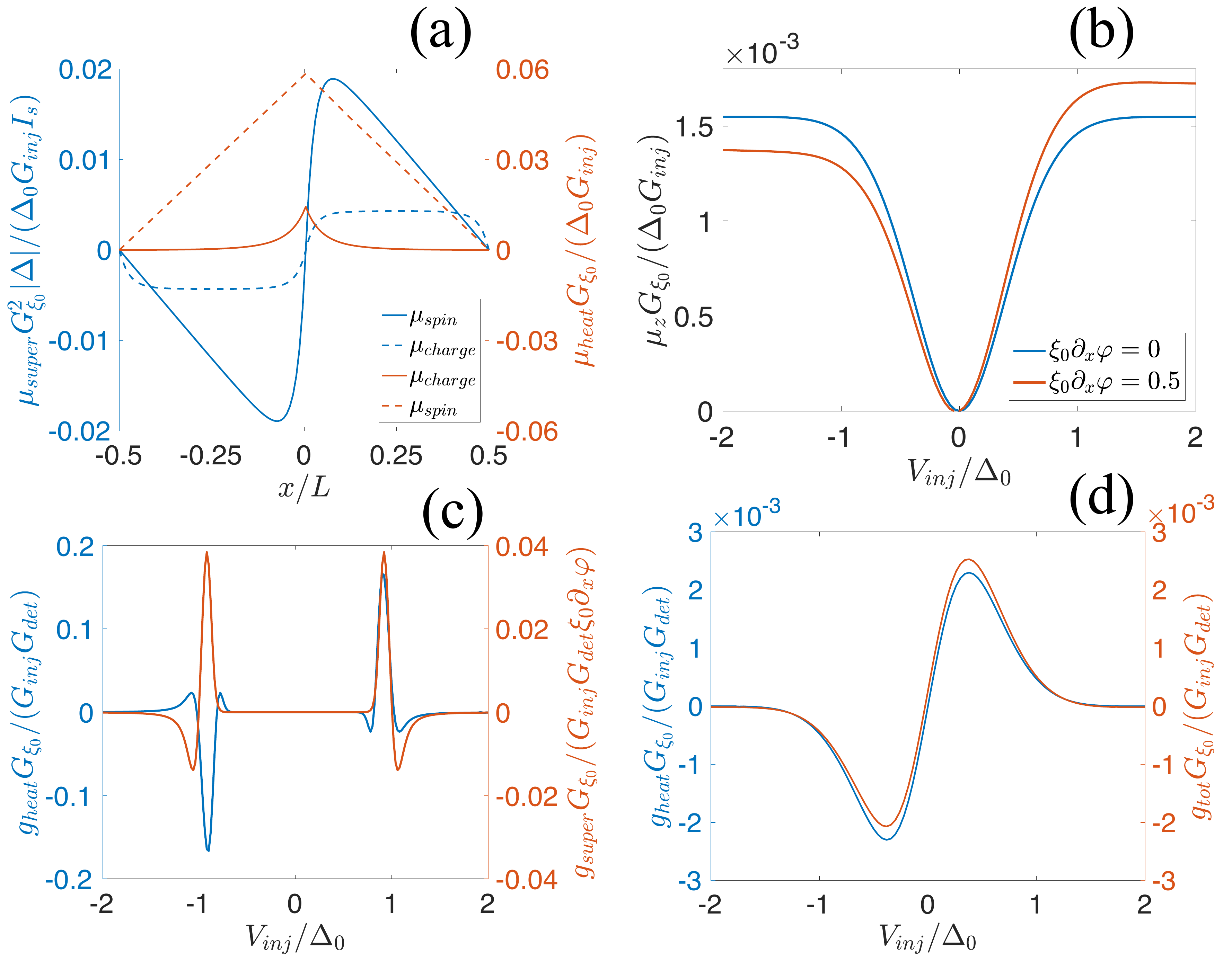}
 \caption{Spin accumulation and nonlocal conductance. (a) Position dependence of heat (red) and supercurrent (blue) induced charge and spin imbalances. Here the results are calculated for $T=0.15\Delta_0$, $h=0.3\Delta_0$ at $V_{\rm inj}=0.1\Delta_0$. The thick curves are odd and dashed curves are even in injection voltage. (b) Injection voltage dependence on spin accumulation for $T=0.25\Delta_0$. (c) Nonlocal conductance as a function of injection voltage in separate scales for heat and supercurrent induced effects (with $\xi_0\partial_x\varphi$=0.1) for $T=0.02\Delta_0$. (d) Heat induced and the total conductance as a function of injection voltage for $T=0.25\Delta_0$ (with $\xi_0\partial_x\varphi$=0.5). The parameters $\tau_{so}^{-1}=0.0475\Delta_0$, $\tau_{sf}^{-1}=0.0025\Delta_0$,  $h=0.05\Delta_0$, and $L=20\xi_0$ are common in panels (b)-(d).}
 \label{fig:figure3}
\end{figure}
For pure spin-flip relaxation, contribution of $f_L$ increases as a
function of the spin relaxation rate, and diverges when the strong relaxation
completely kills superconductivity. This can be seen in the relaxation
rate dependence of $\mu_z$ in the linear response regime in
Fig.~\ref{fig:figure2}b. For pure spin-orbit relaxation, the effect of
the exchange field is suppressed, and thereby also the charge-spin conversion. 

\section{Spin accumulation and nonlocal conductance}
The charge-spin conversion can be detected by inspecting the non-local
conductance $g_{nl}=dI_{\rm det}/dV_{\rm inj}$ in the presence of the
supercurrent $I_s$ driven across the wire. Without supercurrent, this
quantity was measured in Refs.~
\onlinecite{quay2013spin,Hubler2012,PhysRevB.87.024517}. We show an example
of $g_{nl}$ in Fig.~\ref{fig:figure3}c-d. We separate it in
different symmetry components vs.~$V_{\rm inj}$ and $P_{\rm det}$ as
\begin{equation}
\label{eq:nlg}
g_{nl}=g_{ee}+g_{eo}+(g_{oe}+g_{oo})P_{\rm det},
\end{equation}
where $g_{\alpha e/o}(V_{\rm inj}) = \pm g_{\alpha e/o}(-V_{\rm inj})$
and $\alpha=e/o$ describe the symmetry vs.~$P_{\rm det}$. 
Since the derivative of the detector current with respect to $V_{\rm inj}$ flips the parity of the terms, the
conductance due to the pure charge imbalance is even in both $V_{\rm
  inj}$ and $P_{\rm det}$ and hence is described by $g_{ee}$. The term
$g_{oo}=g_{\rm heat}$ is the long-range spin accumulation due to the heat injection 
\cite{silaev2015long,bobkova2015}. The supercurrent induces the term $g_{eo}$ that
describes the conversion of temperature gradients to charge \cite{schmid1979,pethick1980charge,clarke1979supercurrent}, whereas $g_{oe}=g_{\rm super}$ results from the
supercurrent-induced charge-spin conversion. The symmetry of $g_{\rm
  super}$ results from the fact that it is related to spin
imbalance (and therefore antisymmetric in $P_{\rm det}$) and
originates from induced charge imbalance. In normal-metal spin
injection experiments \cite{jedema2001electrical} only the term
$g_{oe}$ is non-zero, but it requires non-zero spin
polarization $P_{\rm inj}$ of the injector. Here $P_{\rm inj}=0$.

The term $g_{\rm super}$ should be compared to the contribution determined by effective heating\cite{silaev2015long} (\ref{Eq:LongRange2})
\begin{equation}
g_{\rm heat} = \frac{G_{\rm inj}}{G_{\xi_0}} \frac{L}{2\xi_0} u_3(x) \int_0^\infty
d\epsilon \frac{\partial n_+}{\partial V_{\rm inj}} \frac{N_\uparrow^2-N_\downarrow^2}{D_L},
\end{equation}
where $u_3(x)=u_3(-x)$ is a linear function interpolating from unity
at the injector to zero at the reservoirs and
$n_+=(n_0(\epsilon+eV)+n_0(\epsilon-eV)-2n_0)/2$. For $T\rightarrow
0$, $\partial n_{\pm}/\partial V_{\rm inj}$ approaches a $\delta$-function at $\epsilon = \pm eV$, and we can estimate the
integrals by the values of the kinetic coefficients at those
energies. For $eV \approx \Delta\pm h$ where the main signal resides,
$g_{\rm super} \approx 2\xi_0 g_{\rm heat}/L$ for $\xi_0 \nabla \varphi \approx
\tau_{\rm orb}^{-1}\Delta+\tau_{\rm sf}^{-1} + \tau_{\rm so}^{-1}$, i.e., when the supercurrent starts affecting the
density of states. At higher temperatures and lower voltages $eV \lesssim k_B
T$, where quasiparticle effects are
visible even at linear response, $g_{\rm super}$ can dominate over
$g_{\rm heat}$.

\section{Conclusion}
In conclusion, we have shown how the nonequilibrium supercurrent in a
spin-split superconductor can partially convert charge imbalance to
spin imbalance. The resulting spin imbalance is long-ranged, 
decaying only due to inelastic scattering. Here we have concentrated on
a setup with
collinear magnetizations.
We expect that the generalization of our theory to the
case with inhomogeneous magnetization would shed light on the possible
coherently controllable nonequilibrium spin torques. We also expect to find analogous effects in superconducting
proximity structures in the presence of spin splitting, i.e.,
combining the phenomena discussed in Refs.~\onlinecite{virtanen04} and \onlinecite{Machon2013}.

\acknowledgments
We thank Manuel Houzet and Marco Aprili for the question that started
this project and  Timo Hyart and Charis Quay for illuminating discussions. This work was supported by the Academy of Finland Center of Excellence (Project No. 284594), Research Fellow (Project No. 297439) and Key Funding (Project No. 305256) programs. 

\appendix

\section{\label{sec:selconeq}Self-consistency equation the for $\Delta$}
The pair potential $\Delta$ should be obtained self-consistently from
\begin{equation}
\Delta=\frac{\lambda}{16}\int_{-\Omega_D}^{\Omega_D}d\epsilon \Tr\bigg[(\tau_1-i\tau_2)\hat{g}^K(\epsilon)\bigg],
\end{equation}
where $\lambda$ is the coupling constant and $\Omega_D$ is the Debye cutoff energy. In the presence of both spin splitting and non-equilibrium distribution functions, this goes to the form \cite{bergeret2017nonequilibrium}
\begin{equation}\label{eq:gapeq}
\begin{split}
\Delta=&\frac{\lambda}{2} \int_{-\Omega_D}^{\Omega_D} d\varepsilon \left[{\rm
  Im}g_{01}^R f_L + {\rm Im} g_{31}^R f_{T3} \right.\\ & \left.+ i ({\rm Re} g_{01}^R
f_T + {\rm Re} g_{31}^R f_{L3})\right],
\end{split}
\end{equation}
where $g_{ij}^R$ is the part of the Retarded Green's function proportional to $\sigma_i \tau_j$. The results obtained in the main text use the self-consistent equilibrium gap, but do not include the nonequilibrium corrections. For the gap amplitude $|\Delta|$ this approximation is justified in the case of low injection conductance $G_{\rm inj}$. However, with such a choice the charge current is strictly speaking not conserved in the presence of a constant phase gradient. This is because the quasiparticle injection modifies the phase of $\Delta$ (the two last terms in Eq.~(\ref{eq:gapeq})), and the true phase gradient corresponding to a constant charge current becomes position dependent. Such an effect is of a higher order in the phase gradient, and within a perturbation approach can therefore be disregarded. We leave such higher-order effects for further work.

\section{\label{sec:KineticCoefficients}Kinetic coefficients}

The Green's function in Eq.~(2) satisfies the normalization condition $\check{g}^2=1$, which allows us to parameterize the Keldysh Green's function as 
$\check{g}^K=\check{g}^R\check{f}-\check{f}\check{g}^A$, where the distribution matrix $\check{f}=f_L+f_T\tau_3+f_{T3}\sigma_3+f_{L3}\sigma_3\tau_3$. We also can parameterize the retarded Green's function as $\check{g}^R=g_{01}\tau_1+g_{02}\tau_2+g_{03}\tau_3+g_{31}\sigma_3\tau_1+g_{32}\sigma_{3}\tau_2+g_{33}\sigma_{3}\tau_3$, and $\check{g}^A=-\tau_3\check{g}^{R\dagger}\tau_3$. Here $g_i$ are complex scalar functions. From these, we identify $N_+=\text{Re}(g_{03})$ and $N_-=\text{Re}(g_{33})$.

The kinetic coefficients $D_i$, $R_i$, and $S_i$ in Eq.~(3) and Eq.~(4) can be expressed in terms of the parameterized functions $\check{g}^R$ and $\check{g}^A$. The $D_i$s are
\begin{align*}
\begin{split}
D_L=&\frac{D}{2}(1-|g_{01}|^2-|g_{02}|^2+|g_{03}|^2-|g_{31}|^2\\&-|g_{32}|^2+|g_{33}|^2)
\end{split}\\
\begin{split}
D_{T3}=&-D\left[ \text{Re}(g_{01}g_{31}^*)+\text{Re}(g_{02}g_{32}^*)-\text{Re}(g_{03}g_{33}^*)\right]
\end{split}\\
\begin{split}
D_T=&\frac{D}{2}(1+|g_{01}|^2+|g_{02}|^2+|g_{03}|^2+|g_{31}|^2\\&+|g_{32}|^2+|g_{33}|^2)
\end{split}\\
\begin{split}
D_{L3}=&D\left[ \text{Re}(g_{01}g_{31}^*)+\text{Re}(g_{02}g_{32}^*)+\text{Re}(g_{03}g_{33}^*)\right].
\end{split}
\end{align*}
The $R_i$s are
\begin{align*}
R_T&=\text{Re}(g_{01})\Delta\cos\varphi-\text{Re}(g_{02})\Delta\sin\varphi\\
R_{L3}&=\text{Re}(g_{31})\Delta\cos\varphi-\text{Re}(g_{32})\Delta\sin\varphi.
\end{align*}
The $S_i$s are
\begin{align*}
\begin{split}
S_{L3}&=\tau_{sn}^{-1}\left\lbrace \text{Re}(g_{03})^2-\text{Re}(g_{33})^2\right.\\
&\left.+\beta\left[\text{Im}(g_{01})^2-\text{Im}(g_{31})^2+\text{Im}(g_{02})^2-\text{Im}(g_{32})^2 \right]   \right\rbrace 
\end{split}\\
\begin{split}
S_{T3}&=\tau_{sn}^{-1}\left\lbrace \text{Re}(g_{03})^2-\text{Re}(g_{33})^2\right.\\
&\left.+\beta\left[\text{Re}(g_{31})^2-\text{Re}(g_{01})^2+\text{Re}(g_{32})^2-\text{Re}(g_{02})^2 \right]   \right\rbrace ,
\end{split}
\end{align*}
where $\tau_{sn}^{-1}=\tau_{so}^{-1}+\tau_{sf}^{-1}$ and the parameter $\beta=(\tau_{so}-\tau_{sf})/(\tau_{so}+\tau_{sf})$ describes the relative strength of the spin-orbit and spin-flip scattering. For $\beta>0$, spin-flip scattering dominates the spin-orbit scattering, and vice versa for $\beta<0$. These coefficients are independent of $\varphi$ (the dependence of $\varphi$ in $R_i$ terms is canceled {by the corresponding terms in} $g_{i}$).

There are also two more coefficients in Eq.~(3) and Eq.~(4), spectral
supercurrent and spectral spin supercurrent, which depend on the phase gradient $\partial_x\varphi$
\begin{align*}
j_E\partial_x \varphi &=\frac{1}{8}D\Tr\left[\left(\check{g}^R\partial_x\check{g}^R-\check{g}^A\partial_x\check{g}^A \right) \tau_3 \right]\\
j_{Es}\partial_x\varphi &=\frac{1}{8}D\Tr\left[\left(\check{g}^R\partial_x\check{g}^R-\check{g}^A\partial_x\check{g}^A \right) \sigma_3\tau_3 \right].
\end{align*}
These two terms are related to the nonzero charge supercurrrent and spin-energy current. Here and below we assume that the wire is in the $x$ direction and all changes in the phase $\varphi$ and the distribution functions take place in that direction. 

\begin{widetext}
\section{\label{sec:PerTheo}Perturbation theory solutions of kinetic equations in the linear order by $\xi_0\nabla\varphi$.}
The general solution of the kinetic equations in Eq.~(3) can be written as
\begin{equation}\label{eq:GeneralSolution}
\begin{pmatrix} f_L,&f_{T3},&f_T,&f_{L3} \end{pmatrix}^{T}=(C_{01}+C_{02}x)\textbf{v}_0^T+C_1e^{k_Lx}\textbf{v}_1^T+C_2e^{-k_Lx}\textbf{v}_2^T+C_3e^{k_{T1}x}\textbf{v}_3^T+C_4e^{-k_{T1}x}\textbf{v}_4^T+C_5e^{k_{T2}x}\textbf{v}_5^T+C_6e^{-k_{T2}x}\textbf{v}_6^T,
\end{equation}
where $\textbf{v}_0^T=(1,0,0,0)^T$, $k_L$, $k_{T1}$ and $k_{T2}$ are the energy dependent inverse length scales,  the other $\textbf{v}_i^T$s can be determined numerically,
and  $C_i$s can be determined from the boundary conditions \eqref{Eq:bcDistrFunc}. 
For a small phase gradient, we can determine these coefficients analytically. {Below we concentrate in particular on the solutions of the modes related to the supercurrent induced spin imbalance and treat the supercurrent  as a perturbation in the kinetic equations.} In the zeroth order Eq.~(3) decouples into two sets of kinetic equations. First we concentrate on the part odd in the injection voltage, describing charge imbalance. In this case, for a vanishing supercurrent the relevant distribution function components are $f_T$ and $f_{L3}$. We denote their values in the absence of supercurrent by $f_T^0$ and $f_{L3}^0$. On the other hand, the supercurrent couples them to the other two functions $f_L$ and $f_{T3}$ and induces the change $\delta f_L$ and $\delta f_{T3}$ which we calculate to linear order in the phase gradient. 
For $f_T$ and $f_{L3}$, we get the first set of kinetic equations
\begin{equation}
\begin{pmatrix}
D_T & D_{L3}\\D_{L3} & D_T
\end{pmatrix}
\begin{pmatrix}
\partial_x^2 f_{T}^0\\\partial_x^2 f_{L3}^0
\end{pmatrix}=
\begin{pmatrix}
R_T & R_{L3}\\R_{L3} & R_T+S_{L3}
\end{pmatrix}
\begin{pmatrix}
f_{T}^0\\f_{L3}^0
\end{pmatrix}. 
\end{equation}
In what follows, we choose $\Delta_0$ as the reference energy scale, and therefore the coherence length $\xi_0=\sqrt{\hbar D/\Delta_0}$ becomes the reference length scale. That means, for example, that the dimensionless quantities describing spin relaxation are of the form $\tau_{\rm sf} \Delta_0$ and $\tau_{\rm so} \Delta_0$. 

Using the boundary conditions \eqref{Eq:bcDistrFunc}, we obtain for $\kappa_I L \ll 1$
\begin{equation}\label{eq:SolutionOffTfL3}
\begin{pmatrix}
f_{T}^0\\f_{L3}^0
\end{pmatrix}
=\kappa_I\xi_0 n_-(\epsilon,V_{inj})\sum_{i=1,2} A_i e^{-k_{Ti}x/\xi_0}
\begin{pmatrix}
k_{R_i}\\-1
\end{pmatrix},\ 0\leq x\leq \frac{L}{2}
\end{equation}
where the inverse length scales
$$
k_{T1/2}^2=\frac{D_T(2R_T-S_{L3})-2D_{L3}R_{L3}{\pm}\sqrt{4(D_TR_{L3}-D_{L3}R_T)^2+4D_{L3}(-D_TR_{L3}+D_{L3}R_T)S_{L3}+D_T^2S_{L3}^2}}{2(D_T^2-D_{L3}^2)},
$$
and the coefficients
$$
A_i=\frac{\left[N_-(D_{L3}-D_Tk_{Ri'})-N_+(D_T-D_{L3}k_{Ri'}) \right]}{4(D_{L3}^2-D_T^2)(k_{Ri}-k_{Ri'})k_{Ti}},
$$
$$
k_{R1/2}=\frac{D_TS_{L3}{\mp}\sqrt{4D_{L3}^2R_T(R_T+S_{L3})-4D_{L3}D_TR_{L3}(2R_T+S_{L3})+D_T^2(4R_{L3}^2+S_{L3}^2)}}{2(D_TR_{L3}-D_{L3}R_T)}.
$$
For the perturbed terms of $f_L$ and $f_{T3}$, we get another set of kinetic equations
\begin{equation}
\begin{pmatrix}
D_L & D_{T3}\\D_{T3} & D_L
\end{pmatrix} 
\begin{pmatrix}
\partial_x^2\delta f_L\\\partial_x^2\delta f_{T3}
\end{pmatrix}+
\begin{pmatrix}
j_E\partial_x\varphi &j_{Es}\partial_x \varphi\\
j_{Es}\partial_x\varphi &j_{E}\partial_x \varphi
\end{pmatrix}
\begin{pmatrix}
\partial_x f_T^0\\\partial_x f_{L3}^0
\end{pmatrix}=
\begin{pmatrix}
0 & 0\\0 & S_{T3}
\end{pmatrix}
\begin{pmatrix}
\partial_x^2\delta f_L\\\partial_x^2\delta f_{T3}
\end{pmatrix}.
\end{equation}
Using the solution in Eq.~(\ref{eq:SolutionOffTfL3}), we obtain 
$$
\begin{pmatrix}
\delta f_L\\ \delta f_{T3}
\end{pmatrix}=
\kappa_I\xi_0^2\partial_x\varphi n_-(\epsilon,V_{inj})\sum_{i=1,2}\left[ \frac{\alpha_i}{k_L^2-k_{Ti}^2} (e^{-k_{Ti}x/\xi_0}-e^{-k_L x/\xi_0})\begin{pmatrix}
-D_{T3}/D_L\\ 1
\end{pmatrix}\right.
$$
\begin{equation}\label{eq:SolutionOffLfT3}
\left.+\frac{\beta_i}{k_{Ti}^2}\left( \frac{2x}{L}-1+e^{-k_{Ti}x/\xi_0}\right)\left(\begin{array}{c}
1\\
0
\end{array} \right)\right],\ 0\leq x\leq \frac{L}{2},
\end{equation}
where the inverse length scale 
$$
k_L^2=\frac{S_{T3}D_L}{D_L^2-D_{T3}^2},
$$
and the coefficients
$$
\alpha_i=\frac{[j_{Es}(D_{T3}+D_{L}k_{Ri})-j_{E}(D_{L}+D_{T3}k_{Ri})][N_-(D_{L3}-D_{T}k_{Ri'})-N_+(D_{T}-D_{L3}k_{Ri'})]}{2(D_T^2-D_{L3}^2)(D_L^2-D_{T3}^2)(k_{Ri}-k_{Ri'})},
$$
$$
\beta_i=\frac{(j_Ek_{Ri}-j_{Es})[N_+(D_T-D_{L3}k_{Ri'})-N_-(D_{L3}-D_Tk_{Ri'})]}{2 D_L(D_T^2-D_{L3}^2)(k_{Ri}-k_{Ri'})}.
$$

The spin accumulation generated from the supercurrent is
$$
\mu_z=\frac{1}{2}\int_0^{\infty}d\epsilon(N_+\delta f_{T3}+N_-\delta f_L)
$$
$$
=\frac{1}{2}\kappa_I\xi_0^2\partial_x \varphi \int_0^{\infty} d\epsilon\ n_-(\epsilon,V_{inj})\sum_{i=1,2}\left[\left( N_+-N_-\frac{D_{T3}}{D_L}\right)\frac{\alpha_i}{k_L^2-k_{Ti}^2} (e^{-k_{Ti}x/\xi_0}-e^{-k_L x/\xi_0}) \right. 
$$
\begin{equation}
\left. +N_-\frac{\beta_i}{k_{Ti}^2}\left(\frac{2x}{L}-1+e^{-k_{Ti}x/\xi_0} \right) \right] ,\ 0\leq x\leq \frac{L}{2}.
\end{equation}

In the extreme limit of $\tau_{sn}^{-1}\rightarrow0$, this result can be reduced to a simpler form. In this case, $S_{T3}$ and $S_{L3}$ terms in the kinetic equations are zero, therefore, $e^{-k_Lx/\xi_0}$ term is replaced by a linear term with same coefficients with $\delta f_L$. For the linear response regime $n_-(\epsilon,V_{inj})=V_{inj}\partial n_0/\partial\epsilon$, we get
$$
\mu_z=V_{inj}\kappa_I\xi_0^2\partial_x\varphi\int_{0}^{\infty}d\epsilon\frac{\partial n_0}{\partial\epsilon}\left[\frac{N_{\uparrow}^2j_s^{\uparrow}}{4D_L^{\uparrow}R_T^{\uparrow}}\left(\frac{2x}{L}-1+e^{-\sqrt{R_T^{\uparrow}/D_T^{\uparrow}}x/\xi_0} \right)\right.
$$
\begin{equation}\label{eq:SpAccForZeroRela}
\left.-\frac{N_{\downarrow}^2j_s^{\downarrow}}{4D_L^{\downarrow}R_T^{\downarrow}}\left(\frac{2x}{L}-1+e^{-\sqrt{R_T^{\downarrow}/D_T^{\downarrow}}x/\xi_0} \right) \right],\ 0\leq x\leq \frac{L}{2},
\end{equation}
where the $\uparrow$ and $\downarrow$ quantities are the addition and subtraction of the singlet and triplet components of the spectral quantities, $j_s^{\uparrow/\downarrow}=j_E\pm j_{Es}$, $N_{\uparrow/\downarrow}=N_+ \pm N_-$,
$D_L^{\uparrow/\downarrow}=D_L \pm D_{T3}$, and
$R_{\uparrow/\downarrow}=R_T\pm R_{L3}$.
  
It is straightforward to see that $\mu_z=0$ for $h=0$, since the quantity $N^2j_s/(D_LR_T)$ is equal for both spin species. For nonzero $h$ the difference of this quantity for different spin species gives the spin accumulation. However, without relaxation, this quantity is proportional to $1/\sqrt{\Gamma}$, which describes the broadening of the spectral quantities. 

In practice, the relevant broadening renormalizing $\mu_z$ comes from the orbital effect due to either a magnetic field or the phase gradient itself \cite{deGennes:566105,belzig96,anthore2003}, or due to terms contributing to the spin relaxation \cite{bergeret2017nonequilibrium}. The two first effects can be described by an orbital relaxation rate $\tau_{\rm orb}^{-1} =(\xi_0 \partial_x \varphi)^2/2+(De^2 B^2 d^2/6)$ \cite{anthore2003}, where $B$ is the magnetic field, and $d$ is the film thickness. In the presence of spin relaxation described by the rate $\tau_{\rm sn}^{-1}$, an estimate for the overall broadening comes from $\Gamma \mapsto \tau_{\rm orb}^{-1} + \tau_{\rm sn}^{-1}$, but the exact amount depends on the relaxation mechanism and the size of the exchange field. As an example, we show the supercurrent induced $\mu_z$ vs. $\tau_{\rm orb}^{-1}$ in Fig.~\ref{fig:figures}(a).
Since $\mu_z \propto (\xi_0 \partial_x \varphi) \Gamma^{-1/2}$, for large phase gradients satisfying $\xi_0 \partial_x \varphi \gg \sqrt{De^2 B^2 d^2/6} + \tau_{sn}^{-1}$, the spin accumulation becomes independent of $\partial_x \varphi$.
\begin{figure}[h!] 
\centerline{
$
\begin{array}{c}
\includegraphics[width=1.0\linewidth]{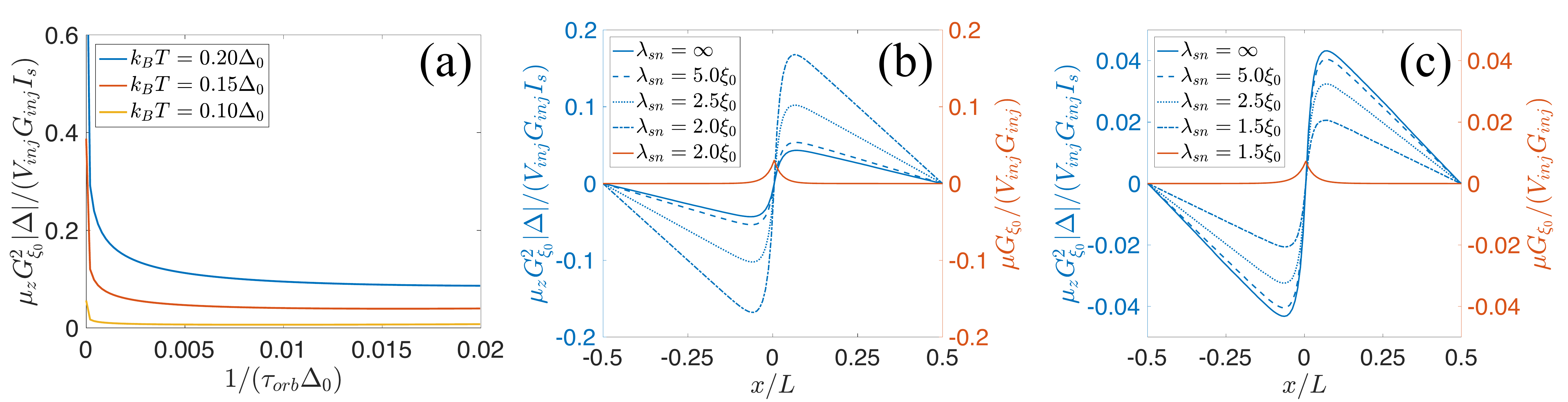}
\end{array}$} 
\caption{  Spin accumulation with and without relaxation in the linear response regime. (a) The dependence on orbital depairing rate in the case without spin relaxation. Position dependence in the case of pure spin-flip relaxation (b) and pure spin-orbit relaxation (c). An exchange field $h=0.3\Delta_0$ is common for all panels, and a temperature $T=0.15\Delta_0$ is used in (b) and (c) panels. The red curves describe the charge imbalance. The spin relaxation length is defined as $\lambda_{sn}=\sqrt{\tau_{sn}D}$.}
\label{fig:figures}
\end{figure}

However, spin relaxation affects also the decay of the nonequilbrium components of the distrubution function via the relaxation terms $\sim S_{T/L3}$. In another extreme limit $\tau_{sn}\rightarrow\infty$, we also can have a simpler form of Eq.~(\ref{eq:SpAccForZeroRela}). In this case $4D_{L3}(D_TR_{L3}-D_{L3}R_T)/D_T^2\ll S_{L3}$, and
\begin{equation}\label{eq:muzInStroRelaxb}
\mu_z=V_{inj}\kappa_I\xi_0^2\partial_x\varphi
\int_0^{\infty}d\epsilon\frac{\partial n_0}{\partial \epsilon}\frac{\left(N_{\uparrow}^2-N_{\downarrow}^2\right)j_E}{4R_TD_L}\left(\frac{2x}{L}-1+2e^{-k_{T2}x/\xi_0}-e^{-k_Lx/\xi_0}\right),\ 0\leq x\leq \frac{L}{2}.
\end{equation}
Here, except the density of states, the triplet component of other spectral quantities do not contribute to the spin accumulation. The difference of the density of states for two spin species behaves differently for spin-orbit and spin-flip relaxations. Spin-orbit relaxation does not affect the pair potential but tries to lift the effect of the spin-splitting field. Therefore, $\mu_z$ approaches zero for very strong relaxation (S4(c)). In the case of spin-flip relaxation, it suppresses the pair potential, therefore, spin-accumulation diverges the strong spin-flip relaxations destroys the superconductivity (S4(b)).
\end{widetext}

\bibliography{./refs}

\end{document}